\documentclass[twocolumn,pra,aps,showpacs]{revtex4-2}
\usepackage{graphicx,graphics}
\usepackage{amsmath}
\usepackage{amssymb}
\usepackage{epstopdf}
\usepackage{amstext}
\usepackage{amsthm}
\usepackage{amsfonts}
\usepackage{latexsym}
\usepackage{array}
\usepackage{bm}
\usepackage{xfrac}
\usepackage{lipsum}
\usepackage[toc,page]{appendix}
\usepackage{subfigure}
\newcommand{\igbjd}[1]{}\newcommand{\beqa}{\begin{eqnarray}}
\newcommand{\eeqa}{\end{eqnarray}}
\newcommand{\beq}{\begin{equation}}
\newcommand{\eeq}{\end{equation}}
\usepackage[normalem]{ulem}
\usepackage[colorlinks=true,citecolor=Cerulean,linkcolor=RubineRed,urlcolor=Cerulean,pdftex]{hyperref}
\usepackage{xcolor}
\definecolor{Cerulean}{rgb}{0.,0.59,0.835}
\definecolor{RubineRed}{rgb}{0.61,0.07,0.12}

\begin{document}
\title{Off-diagonal long-range order in arrays of dipolar droplets}
\author{R. Bomb\'in}
\affiliation{Departament de F\'isica, Campus Nord B4-B5, 
Universitat Polit\`ecnica de Catalunya,
E-08034 Barcelona, Spain}
\author{F. Mazzanti}
\affiliation{Departament de F\'isica, Campus Nord B4-B5, 
	Universitat Polit\`ecnica de Catalunya,
	E-08034 Barcelona, Spain}
\author{J. Boronat}
\email{jordi.boronat@upc.edu}
\affiliation{Departament de F\'isica, Campus Nord B4-B5, 
	Universitat Polit\`ecnica de Catalunya,
	E-08034 Barcelona, Spain}

\date{ \today}

\begin{abstract}
We report quantum Monte Carlo results of harmonically confined quantum Bose
dipoles within a range of interactions covering the evolution from a gas
phase to the formation of an array of droplets. Scaling the experimental
setup to a computationally accessible domain we characterize that evolution
in qualitative agreement with experiments. Our microscopic approach generates
ground-state results free from approximations, albeit with some controlled
statistical noise. The simultaneous estimation of the static structure factor
and the one-body density matrix allows for a better knowledge of the quantum
coherence between droplets. Our results show a narrow
window of interaction strengths where diagonal and off-diagonal long-range 
order can
coexist. This domain, which is the key signal of a supersolid state, is
reduced with respect to the one predicted
by the extended Gross-Pitaevskii
equation. Differences are probably due to an increase of attraction in our
model, observed previously in the calculation of critical atom numbers for
single dipolar drops.

\end{abstract}

\maketitle

\section{Introduction}

The experimental realization of cold Bose and Fermi gases with dipolar
interactions opened a very rich landscape to explore new phenomena~\cite
{chrom, chromfermi,lu2011,lu2012, Aikawa2012}. What makes these systems
exceptional and motivates the interest, from theory and experiment, is the
interplay between a long range interaction and their anisotropic
character~\cite{Chomaz2023}. The first experimental realization of a dipolar
quantum gas was carried out with $^{52}$Cr~\cite{chrom} atoms, which have a
magnetic dipolar moment $\mu=6 \,\mu_B$ with $\mu_B$ the Bohr magneton. By
bringing the system close to a Feshbach resonance, it was possible to
significantly reduce the scattering length, making it much smaller than the
dipolar length and thus enhancing its dipolar character. More recently, the
achievement of Bose-Einstein condensate (BEC) states of lanthanides Er~\cite
{Aikawa2012} and Dy~\cite{lu2011,lu2012} atoms, with larger magnetic moments,
$\mu=7 \,\mu_B$ and $10 \,\mu_B$, respectively, has given access to a regime
where the dipolar interaction dominates. 
This is more evident by comparing the 
dipolar length of the three elements $a_{\text{dd}}=15 \,a_0$, $65.5 \,a_0$, 
and 
$131 \,a_0$ ($a_0$ is the Bohr radius) for Cr, Er, and Dy, respectively. 

The dipolar interaction between head-to-tail oriented dipoles is attractive
while it becomes repulsive in parallel configurations.  In the absence of any 
other interaction to compensate
the attraction, mean-field theories predict a collapsed state~\cite{Santos1}. 
The scenario is similar to the one of attractive Bose-Bose
mixtures and, in both cases, it has been found that the introduction of
quantum fluctuations through the perturbative Lee-Huang-Yang (LHY) correction
is able to stabilize the system~\cite{Petrov,Lima2011,Lima2012}. The result
is the formation of self-bound liquid droplets of extremely low densities,
much lower than other liquids in Nature~\cite{Kadau2016}. At
difference with spherical Bose-Bose drops, which can be produced without
confinement, dipolar ones are elongated and require of the presence of an 
external potential.

Squeezing the harmonic trap in the direction of the oriented magnetic
moments, and varying the scattering length $a$, different phases have been
observed~\cite{Chomaz2016}. Increasing the
ratio $\epsilon_{\text{dd}}=a_{\text{dd}}/a$, the system evolves from a gas 
phase to the
formation of a single drop, followed by the appearance of an array of aligned
drops. Interestingly, this array is initially coherent and then becomes
insulated, just as in a crystal of droplets. Within the small window
where local ordering and coherence are simultaneously observed, 
this phase constitutes a supersolid of 
droplets~\cite{Bottcher2019,Tanzi2019b,Chomaz2019}. This ordered
superfluid phase arises from the coexistence of the 
droplets with a small density bath that \textit{connects} the droplets. In this 
sense, this is
significantly different from the supersolid phase that has been searched for
many years in hcp solid $^4$He~\cite{Balibar2010}.

The most common theoretical tool used in the study of dipolar droplets, and
its eventual supersolid character, is the extended Gross-Pitaevskii equation
(eGPE), including LHY beyond mean-field 
effects~\cite{Lima2011,Lima2012,Bisset2016,Bisset2021}. The range of
interaction parameters where supersolidity is observed is quite
narrow~\cite{Chomaz2023}, and has been estimated in two different ways. In
the first case, a monopole mode is induced and the dynamic response of the 
system is measured~\cite{Tanzi2019b}. Within the supersolid regime two modes 
appear,
the lowest one being associated to the coherence between
drops. This low-frequency mode is the signature of superfluidity and
disappears above a characteristic value of the scattering 
length~\cite{Tanzi2019b}. In the second approach, the superfluid density is
obtained from the response of the system to a translational 
movement~\cite{Tanzi2021}. 

In the present work, we report results for a harmonically-confined dipolar 
gas of $^{162}$Dy atoms, obtained from an ab initio approach that relies 
exclusively
on the microscopic Hamiltonian of the system. Our results are valid in the
zero temperature limit, as they have been obtained using the path
integral ground-state (PIGS) method~\cite{SarsaPIGS,Rota10}, a 
ground-state version of the path 
integral Monte Carlo (PIMC) method devised for quantum simulations at finite 
temperatures. Since
PIGS solves exactly the many-body boson problem, up to some statistical noise, 
one
has direct access to two-body distributions, which are not accessible using
the eGPE due to its mean-field character. By tuning the $s$-wave scattering
length $a$, which we obtain from the zero-momentum limit of the two-body
scattering T-matrix~\cite{pfau_bombin}, we study the evolution of the
harmonically trapped dipolar gas. We observe the formation of self-bound
droplets and calculate the static structure factor
and the one-body density matrix
as a function of $a$. The simultaneous observation of Bragg peaks in
the former and off-diagonal long-range-order in
the latter proves the presence of a supersolid state. Our results provide a
narrow window of interaction strengths where it is plausible to conclude that
this exotic state of matter is observed.

The paper is organized as follows. The PIGS method and the main
ingredients of our study are discussed in Sec. II. Results are presented and
discussed in Sec. III, paying particular attention to the static structure
factors and one-body density matrix. Finally, the main conclusions are
reported in Sec. IV.

\section{Quantum Monte Carlo method}

The starting point in any ab initio microscopic approach to a quantum many-body 
problem is the Hamiltonian. In the present case, it is given by
\begin{equation}
\hat{H}=-\frac{\hbar^2}{2m}\sum_{i=1}^N\nabla^2_i+
  \sum_{i<j}^{N} (V_{\text d} (\bm{r}_{ij}) + V_{\text{s}} (r_{ij}) ) +
\sum_{i=1}^{N} V_{\text{h}} (\bm{r}_i) \ ,
\label{hamiltonian}
\end{equation}
with $\{\bm{r}_i\}$ the particle position coordinates, and 
\begin{equation}
V_{\text d} (\bm{r}_{ij}) =  
\frac{C_{\text{dd}}}{4\pi} \, \frac{1-3 \cos^2\theta_{ij}}{r_{ij}^3}
\label{vdipolar}
\end{equation} 
the dipolar interaction. In Eq. (\ref{vdipolar}),   $C_{\text{dd}}$ is 
proportional to the square of
the dipolar moments, and $(r_{ij},\theta_{ij})$ stands for the polar
coordinates representation of $\bm{r}_{ij}$. The term 
$V_{\text{s}}({\bf r}_{ij}) = V_{\text{s}}(r_{ij})$ is a central potential
that is repulsive at short distances to prevent collapse. It has been chosen
to be of the 6-12 Lennard-Jones form, as in Ref.~\cite{pfau_bombin}, 
\begin{equation}
V_{\text{s}}(r)  =  \left( \frac{\sigma_{12}}{r} \right)^{12} -
  \left( \frac{C_6}{r^6} \right) \ .
\label{ljones}
\end{equation}
The coefficient $C_6$ in Eq. (\ref{ljones}) is known for
Dysprosium ($C_6=2.978\,10^{-2}$ in dipolar units)~\cite{Li2017}. The second 
parameter 
$\sigma_{12}$ is fixed by imposing that the full 
interaction ($V_{\text{s}}+ V_{\text d}$) has the 
desired $s$-wave scattering length. This is accomplished by solving the low
momentum limit of the scattering T-matrix~\cite{pfau_bombin}. The values of   
$\sigma_{12}$ decrease monotonously with $a$ from $7.852\,10^{-4}$ at $a=71.1$ 
to $2.954\,10^{-4}$ at $a=45$. 
Finally, $V_{\text{h}}(r)$ is a confining harmonic potential. 

In our study, we use the first-principles PIGS algorithm which is 
an exact stochastic projection method used to solve the Bose $N$-body
quantum problem. Basically, it consists in a systematic improvement of a
trial wave function $\Psi_T$ by repeated application of the propagator in
imaginary time, which drives the system to the ground state \cite
{SarsaPIGS}, 
\begin{equation}
 \Psi_{\text{PIGS}}(\bm{R}_M) = \int \prod_{i=1}^{M} d\bm{R}_{i-1} 
 G(\bm{R}_i,\bm{R}_{i-1};\tau) \Psi_T(\bm{R}_0) \ ,
\label{Eq_PIGSwf}
\end{equation}
with $\bm{R}_i = \{ \bm{r}_{i;1},\bm{r}_{i;2}, \ldots ,\bm{r}_{i;N} \}$, 
$G(\bm{R}',\bm{R};\tau)=\langle \bm{R}' \vert e^{-\tau\hat{H}}\vert \bm{R} 
\rangle$ the imaginary-time propagator, and $M$ the number of intermediate 
imaginary-time steps.

While the exact $G(\bm{R}',\bm{R};\tau)$ is in general not accessible,
short-time approximations are available and allow for the evaluation of
averages of diagonal observables, mapping the quantum many-body system onto a
classical system of $N$ interacting \textit{polymers} composed by $2 M +1$
beads. Each one of these terms represents a time slice of the evolution from
the initial trial state $\Psi_T$ to a final wave function 
$\Psi_{\text{PIGS}}$ that is sampled at the central bead. 
By increasing $M$ one is able to reduce the systematic error and therefore to
exactly recover the ground-state properties of the system up to residual 
statistical noise.

A good approximation for the propagator $G(\bm{R}',\bm{R};\tau)$ is capital to 
improve the
numerical efficiency of the method. This greatly reduces the complexity of
the calculation but also helps sorting possible ergodicity related issues,
allowing to simulate the quantum system with few beads and a larger time
step. By using a high-order approximation for the propagator, as the one used
in this work, it is possible to obtain an accurate description of the exact
ground-state wave function, even when the initial trial state 
contains no more information than the bosonic statistics, already present
in $\Psi_T=1$ \cite{Rota10}. The PIGS method has been used previously in the
determination of the critical atom number required for droplet
formation~\cite{pfau_bombin} and in the study of two-dimensional(2D) dipolar
gases~\cite{Macia,Bombin_stripes}. Its extension to finite temperature, the
path integral Monte Carlo method, has also been used to study the BKT phase
transition in 2D dipolar gases~\cite{Bombin_bkt}.

\begin{figure*}[t]
  \centering
  \includegraphics[width=0.9\textwidth,angle=0]{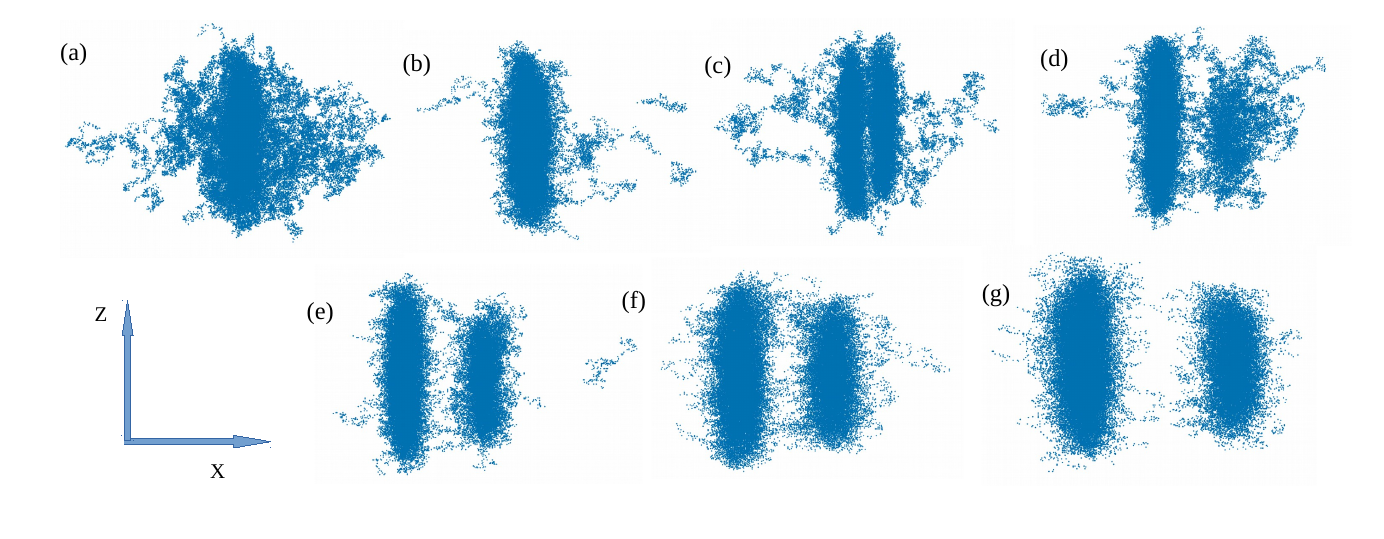}
  \caption{Evolution with the $s$-wave scattering length of the column
   densities of the system along the $y$ direction for $\lambda=1.83$. From
   (a) to (g) $a/a_0=$ 71.1, 60, 58.5, 56.6, 54.9, 51.1, and 45. The size of 
the system in the $x$ direction is $70\,r_0$ for the largest $a$ and reduces 
with decreasing $a$ until $25\,r_0$ for  $a/a_0=45$. In the $z$ direction, it 
extends $50\,r_0$ and squeezes to $40\, r_0$ for the two lowest $a$ values. }
   \label{allI}
\end{figure*}

The presence of diagonal order in the system is studied from the analysis of
the static structure factor $S(\bm{k})$, defined as
\begin{equation}
S(\bm{k}) = \frac{1}{N} \, \langle \hat{\rho}^\dagger (\bm{k}) 
\hat{\rho} (\bm{k})  \rangle   \ ,
\label{sk}
\end{equation}
with $\hat{\rho} (\bm{k}) = \sum_{i=1}^{N} \exp(i \bm{k} \cdot \bm{r}_i)$ the
density fluctuation operator. In PIGS, $S(\bm{k})$ and other diagonal
magnitudes, such as the density profiles $\rho(r)$, are obtained using the
coordinates of the middle points of the chains since it is in these 
points where the exact ground-state wave function is sampled, provided a 
sufficiently large number of beads is used.

One of the main goals of our work is to estimate the possible off-diagonal
long-range order (ODLRO) present in the system, particularly when the formation of
droplets is observed. To this end, we calculate the one-body density matrix (OBDM),
\begin{equation}
\rho_1(\bm{r}_1,\bm{r}^\prime_1) = \frac{\int d \bm{r}_2 \ldots d \bm{r}_N
\Psi^*_0(\bm{R})\Psi_0(\bm{R}')}{\int d\bm{r}_1 \ldots d \bm{r}_N
|\Psi_0(\bm{R})|^2} \ ,
\label{Eq_OBDMPsi}
\end{equation}
where the configuration $\bm{R}$  differs from $\bm{R}'$ only by the position
of one atom. In homogeneous systems, translational invariance makes the OBDM
a function of the coordinate differences only, so $\rho_1({\bf r}_1,
{\bf r}_1') = \rho_1({\bf r}_1 - {\bf r}_1') = \rho_1({\bf r})$. If the
system is anisotropic, as in the current case, $\rho_1$ can be expanded in
partial waves 
\begin{equation}
\rho_1({\bf r}) = \sum_{l,m} \rho_1^{l,m}(r) Y_{l,m}(\theta,\phi) \ ,
\label{rho1_pw}
\end{equation}
with $(r,\theta,\phi)$ the polar coordinates of vector ${\bf r}$ and $Y_
{l,m}$ the Spherical Harmonics. The lowest order mode, corresponding to
$l=m=0$, provides the only isotropic contribution and is the one being
analyzed in this work. In the following, we simplify the notation and denote 
$\rho_1^{0,0}(r)$ as $\rho_1(r)$.
In extended systems, one concludes that ODLRO exists
if a plateau is observed in $\rho_1^{0,0}(r\to\infty)$, which coincides with
the condensate fraction value $n_0$. In PIGS, the expectation value of
non-diagonal observables, like $\rho_1$, is computed mapping the quantum
system into the same classical system of polymers as in the diagonal case,
but cutting one of these polymers in the mid point. 
Building the histogram of
distance frequencies between the cut extremities of the two half polymers,
one can compute the isotropic contribution $\rho_1(r)$. This is effectively 
carried out using the worm algorithm (WA), a technique 
previously developed for path integral Monte Carlo simulations at finite 
temperature \cite{BoninsegniWorm}. The key
aspect of the WA is to work in an extended configuration space, containing
both diagonal (all polymers with the same length) and off-diagonal
(one polymer cut in two separate halves) configurations.

\section{Results}

 We study a system of $^{162}$Dy atoms with a magnetic dipolar
moment $10\,\mu_B$. All our simulations contain $N=400$ harmonically confined
particles. We optimize the number of time steps (beads) and the time-step
itself, resulting in the values $M=30$ and $\tau=0.3$. We use dipolar units
everywhere, with $r_0 = \frac{mC_{\text{dd}}}{4\pi\hbar^2}=3a_
{\text{dd}}$, except when stated otherwise. The experimental setup of 
Ref.~\cite{fau_exp} uses $N=40000$ particles within an harmonic potential of 
oscillator lengths $l_x = 68.908\, r_0$ and $l_y=l_z=37.744\,r0$. Approximating 
the trap by a cylinder  the mean experimental density is 
$\rho \simeq 0.06485\, r_0^{-3}$. In our simulations,  the size of the trap is 
adjusted to reproduce this experimental density. Notice that we do 
not have access to the very large
number of particles used in realistic experimental setups due to the large 
computational cost
involved in our first-principles calculations. Therefore, our scaled system 
serves
to study the physical behavior of the real dipolar gas in a shifted interval
of scattering lengths. We have verified that reducing in our simulations the 
number of particles to $N=200$ we recover the same evolution that the one 
obtained with $N=400$ but reducing adequately the scattering lengths. The 
progressive reduction of $a$, and thus an increase of the interatomic 
attraction, with the size of the system is directly related to the existence of 
a minimum critical atom number for the formation of a self-bound drop.

In the following, we report results obtained with two different confinement
configurations that we characterize by their aspect ratio $\lambda = l_x/l_z$,
with $l_i$ the oscillator lengths along the three spatial directions
($i=x,y,z)$. The first one ($\lambda = 1.83$) corresponds to the scaled
experimental trap~\cite{fau_exp}, with oscillator lengths $l_x=14.84\,r_0$ and 
$l_y=l_z=8.13\,r_0$.
The second one ($\lambda = 5.00$) is more one-dimensional looking, keeping
again the central densities close to the experimental ones: $l_x=29.06\,r_0$ and
$l_y=l_z=5.81\,r_0$.

\begin{figure*}[t]
  \centering
  \includegraphics[width=\textwidth,angle=0]{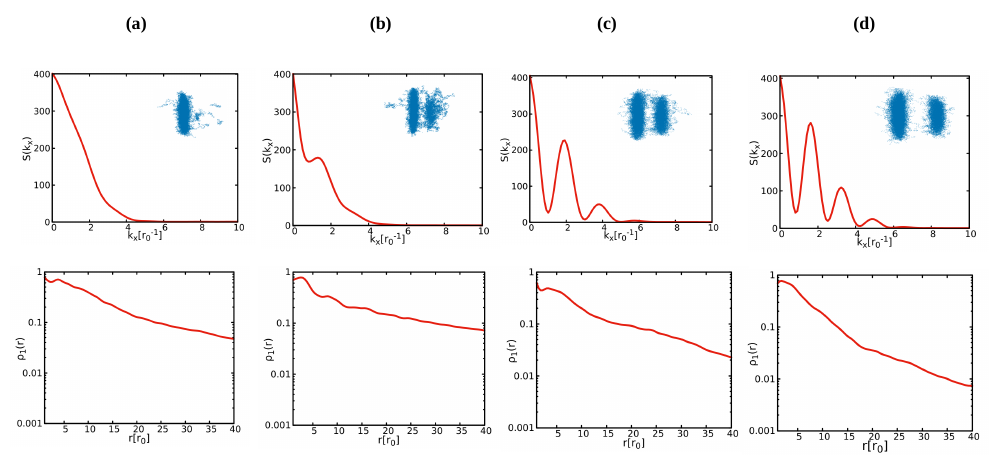}
  \caption{Static structure factor $S(k_x)$ for momenta along the $x$
   direction(top row) and isotropic mode of the one-body density matrix
   $\rho_1(r) $(bottom row) for a set of configurations corresponding
   to $\lambda = 1.83$. From a) to d) the scattering length decreases $a/a_0=
   60$, $56.6$, $51.1$, and $45$. In all cases, the error bars are 
compatible with the width of the lines. The insets show the distribution of
   particles in the $x$-$z$ plane.}
   \label{rhoI}
\end{figure*}

\subsection{Aspect ratio $ \lambda = 1.83$}

First, we study the evolution of the system under the change of the
$s$-wave scattering length $a$ for $\lambda=1.83$. The value of $a$ comes from 
a delicate balance
between the short-range repulsive interaction and the dipolar term. As a
general rule, when $a>0$ decreases, the system becomes less repulsive as the
hard-core radius decreases. In all our simulations we consider the dipoles to
be oriented in the $z$ direction due to the action of an external magnetic
field. Figure~\ref{allI} shows the evolution of the system with decreasing
$a$ by projecting the position of particles in the $x$-$z$ plane. As we are
dealing with a quantum system, particles are delocalized. This feature is
implicit in the path integral method since every particle is represented by a 
chain or
polymer characterized by a given number of beads. The same figure shows
typical configurations once the system has reached equilibrium.

\begin{figure}[tb]
  \centering
  \includegraphics[width=0.8\linewidth,angle=0]{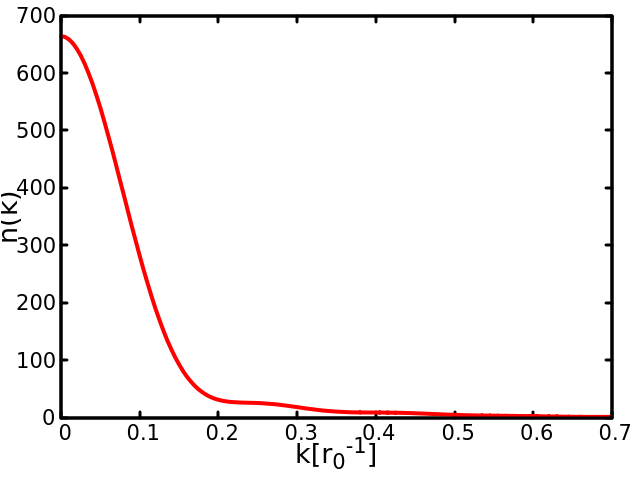}
  \caption{Momentum distribution $n(k)$ for the configuration (b) of Fig. 
\ref{rhoI}.}
    \label{nk}
\end{figure}

Starting from the largest $a=71.1\,a_0$, we observe how the particles tend to
align in the $z$ direction since dipoles are attractive in head-to-tail 
configurations, a  magnetostriction effect~\cite{Stuhler2007}.
This alignment is however not complete
because the gas is harmonically confined, also in that axis. For the largest
$a$ considered, we observe the formation of a single self-bound drop,
surrounded by a gas. Reducing $a$ the system starts to split in two 
drops. For $a=58.5\,a_0$ the
two drops are quite close but tend to separate when the scattering length is
further reduced. Finally, for  $a=45\,a_0$, we observe that the two drops
become more mutually repulsive and are well separated. In the last
configuration, we find two clearly defined drops, with no other particles
around them. The reported evolution with the scattering length reproduces
qualitatively what is actually observed in experiments, once properly scaled
in number of particles and $a$. 

\begin{figure}[tb]
  \centering
  \includegraphics[width=0.9\linewidth,angle=0]{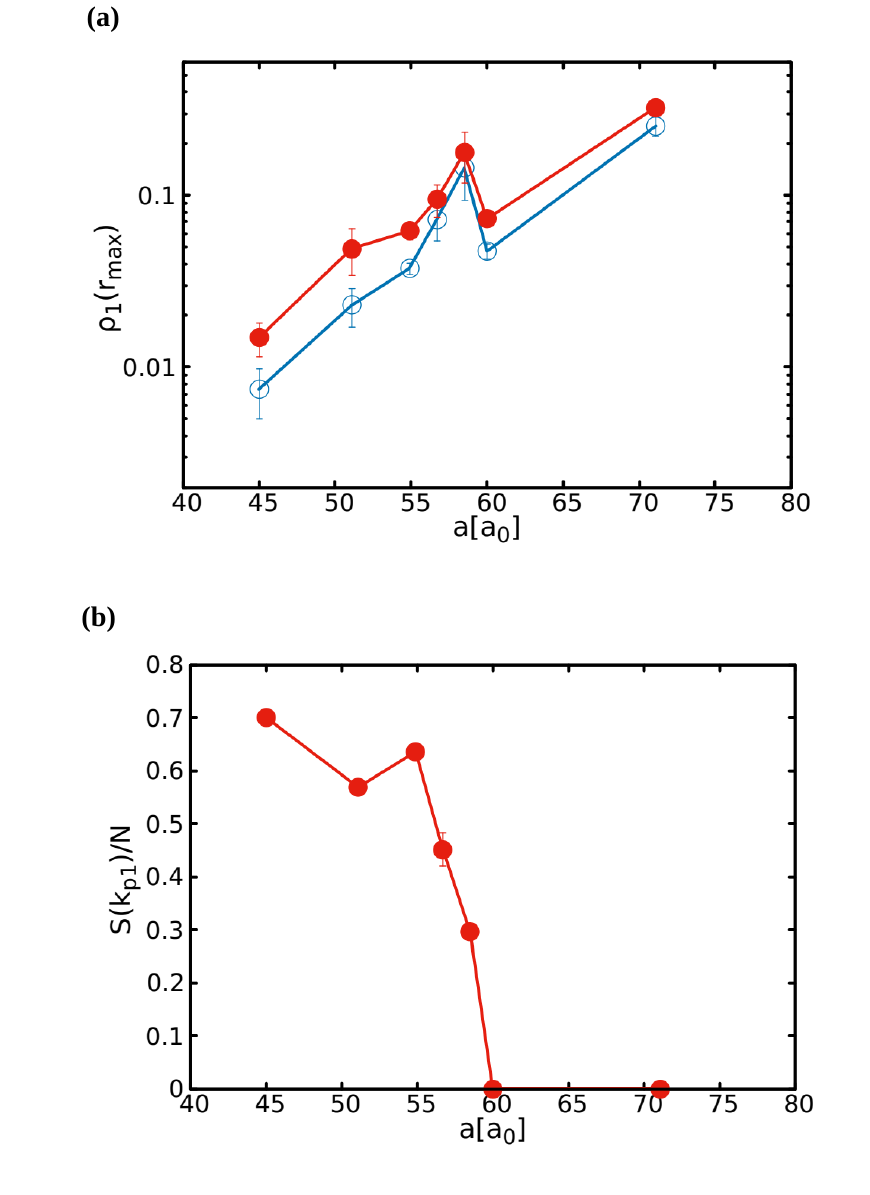}
  \caption{Upper panel: value of $\rho_1(r)$ at a distance $r_{\text max}$ as
   a function of the $s$-wave scattering length $a$. Blue and red points and
   lines correspond to $r_{\text max}=40$ and 30, respectively. Lower panel:
   height of the first peak of $S(k_x)$ as a function of $a$.}
    \label{endpoint}
\end{figure}

\begin{figure*}[t]
  \centering
  \includegraphics[width=0.9\textwidth,angle=0]{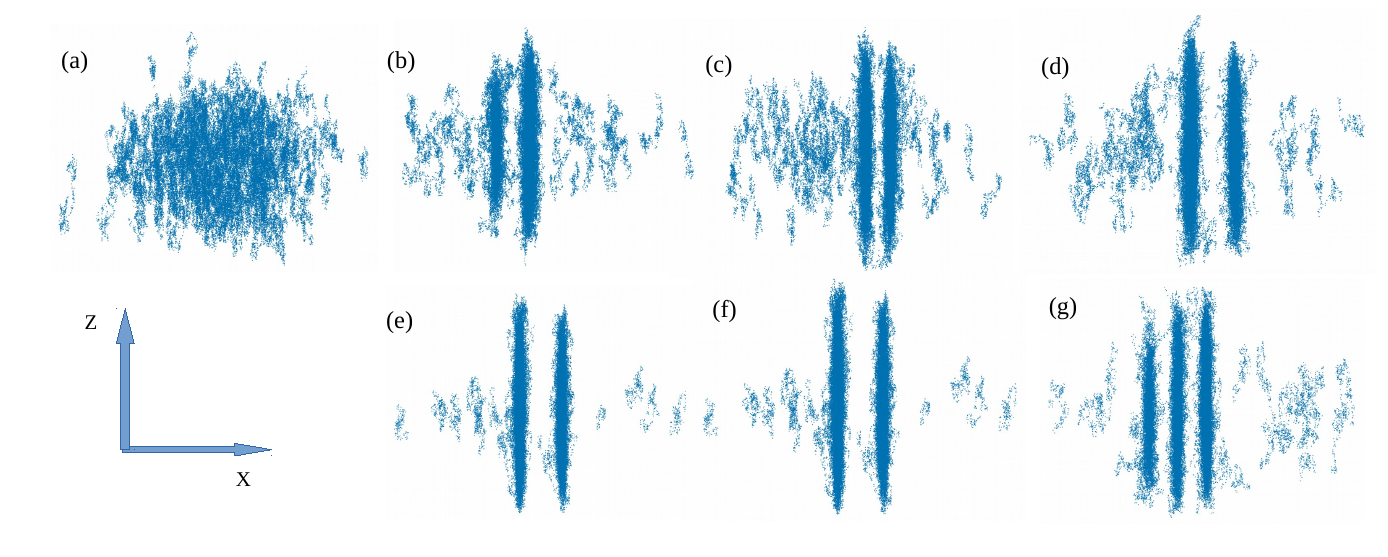}
  \caption{Evolution of the dipolar system with the $s$-wave scattering length
   for $\lambda=5.00$, with the drops projected in the $x$-$z$ plane. From
   (a) to (g) $a/a_0 = $ 71.1, 60, 58.5, 56.6, 54.9, 51.1, and 45.
   The size of 
the system in the $x$ direction is $160\,r_0$ for the largest $a$ and reduces 
with decreasing $a$ until $90\,r_0$ for  $a/a_0=45$. In the $z$ direction, it 
keeps stable around $40\,r_0$.}
  \label{allII}
\end{figure*}

As commented in the preceding section, the use of a microscopic approach such
as the PIGS method allows for the calculation of two of the most relevant
ground-state magnitudes of the system, the static structure factor and the
one-body density matrix. Figure~\ref{rhoI} shows results for both magnitudes
for a selected set of interaction strengths. The static structure factor is
calculated for momenta along the $x$ direction, where drops align. In panel
(a), corresponding to $a=60\,a_0$, a single peak at $k_x=0$ appears since
only a single drop is present. Moving to panels (b) to (d) more peaks
at $k_x>0$ appear in a progressive way, in agreement with the formation of
new drops. It is worth mentioning that the evolution with $a$ of $S
(k_x)$ reproduces qualitatively the experimental data on Dy atoms of
Ref.~\cite{fau_exp}. The peaks in $S(k_x)$ resemble the Bragg peaks of a real
crystal, but they are clearly different since in this case the solid
arrangement is formed by drops rather than by single atoms. 

The bottom panel of Fig.~\ref{rhoI} displays our results for 
$\rho_1(r)$.  
As the system is harmonically confined in all directions,
$\rho_1(r)$ tends to zero when $r$ grows, no matter the direction. The
key point is to check whether this function shows a plateau at intermediate
distances, where the effect of confinement is less relevant. When more than
one droplet appear, this distance is of the order of the separation between 
drops. Configuration (b) containing two drops 
is compatible with the
presence of ODLRO, with a small plateau at $r = 10$ -- $15\,r_0$. 
For this particular interaction, we show the momentum distribution $n(k)$ 
obtained by Fourier transforming $\rho_1(r)$ in Fig.~\ref{nk}. The extent of 
$n(k$ is quite narrow, with a large peak centered al $k=0$. When 
$a$ is further decreased, as in configurations (c) and (d), this plateau is
absent and $\rho_1(r)$ decreases monotonically, so no ODLRO appears.
In this way, therefore, our results seem to indicate that the key ingredients
of supersolidity can be present, but only in a quite narrow window of 
interaction strengths.

To localize the range of scattering length values where a supersolid
behavior could be observed, we show in Fig.~\ref{endpoint} particular values
of both $\rho_1(r)$ and $S(k_x)$. In panel (a), we report the value of
the one-body density matrix for two-values of a large distance $r_
{\text max}$, as a function of $a$. 
Panel (b) shows the value of the first peak at $k_x>0$ of $S(k_x)$, also as a
function of $a$. The strength of the peak in the static structure factor
increases when the drops are more dense, and the amount of gas in between 
decreases. The peak appears when $a < 60\,a_0$ and saturates at $a \simeq
55\,a_0$. Looking at panel (a) one can see that in this same interval $a =
55$ --  $60\,a_0$ the long-distance value of $\rho_1(r)$ is
stable and large. Therefore, it is within this range of interactions where a
superfluid of droplets could exist in our scaled system.

\subsection{Aspect ratio $ \lambda = 5.00$}

The second configuration considered in this work is set to $\lambda=5$,
corresponding to a cigar shape geometry. Compared with the previous case, we 
expect to find more
drops, within the same range of interactions. 
We show the evolution of the system with decreasing $a$ in 
Fig.~\ref{allII}. At the largest value of $a$ considered ($70\,a_0$) the
system looks like a gas, with no trace of drops being formed. However, at
$a=60\,a_0$ two drops, immersed in a cloud of gas, are clearly visible.
Decreasing $a$ even more, the two drops separate and the amount of gas
surrounding them diminishes. Finally, a third drop appears at the lowest
value of $a$ considered. In this case, the separation between them is
noticeably reduced when compared to the configurations that contain only two 
droplets, while there is still a small amount of gas around them.

\begin{figure*}[t]
  \centering
  \includegraphics[width=\textwidth,angle=0]{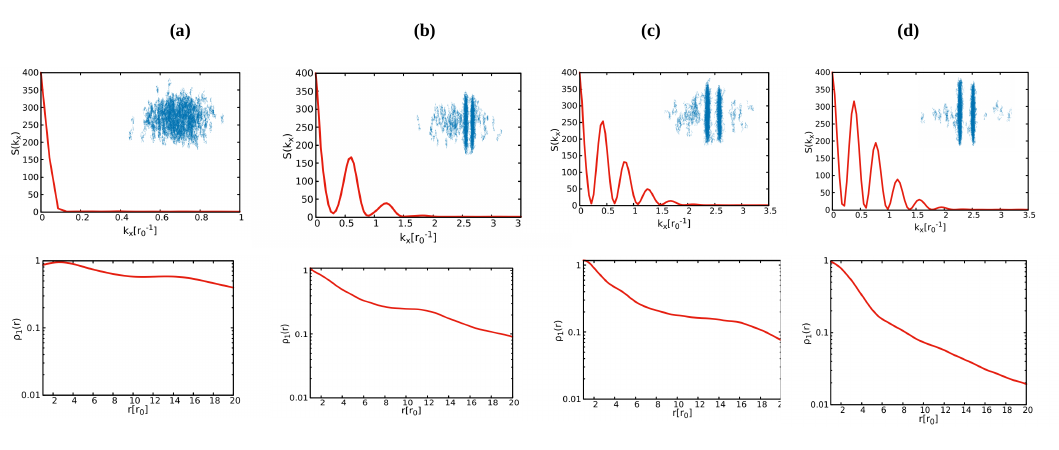}
  \caption{Static structure factor for momenta in the $x$ direction (top row)
   and one-body density matrix $\rho_1(r)$ (bottom row) for a set of
   different configurations with $\lambda=5.00$. From a) to d) the scattering
   length decreases, $a/a_0= 60$, $56.6$, $51.1$, and $45$. 
   In all cases, the error bars are 
compatible with the width of the lines. The insets show
   the distribution of particles in the $x$-$z$ plane.}
    \label{rhoII}
\end{figure*}

Figure~\ref{rhoII} shows results for the static structure factor and the
isotropic component of the one-body density matrix for different interaction
strengths. As it can be seen, the results are qualitatively similar to the ones
found for $\lambda=1.83$. As before, peaks in $S(k_x)$ at $k_x>0$ appear when 
drops are being formed,
and their number and strength increase when the drops are more dense and well
separated. Furthermore, the one-body density matrices decay to zero at
sufficiently long distance but differences between them are appreciated at
intermediate $r$. The emergence of a plateau is clearly visible in panels
(b) and (c). Within the range of interaction strengths between these two
panels, we can state that the system presents simultaneously diagonal and
off-diagonal long-range order, the key signals to supersolidity.

\section{Conclusions}

In the present work we have studied a harmonically trapped gas of 
$^{162}$Dy atoms,
with a large magnetic moment, using a first-principles quantum Monte Carlo
method. The ground-state of this system has been studied previously using the
extended Gross-Pitaevskii equation that incorporates Lee-Huang-Yang
terms~\cite{Lima2011,Lima2012,Bisset2016,Bisset2021}. At difference with the
eGPE approach, our description is fully microscopic, relying directly on the
Hamiltonian of the system and on a projecting method based on the stochastic
evaluation of path integrals~\cite{Rota10}. In order to make the calculations
feasible in practical (computer time) terms, we have scaled the system and
used 400 atoms, keeping the same central density of the trap as in
experiment~\cite{fau_exp}. Having access to the actual position of the atoms
and its delocalization (number of beads in the path integral representation)
our results are somehow closer to experimental views~\cite{Chomaz2023}.

Having access to the atomic coordinates in a spatial representation of the
ground state implies that we can calculate its diagonal and off-diagonal
properties in an accurate way. The PIGS method provides exact estimates of
ground-state properties within some residual statistical noise.  We study the 
evolution of the confined system with $a$, which decreases when the interaction 
becomes more attractive. 

The diagonal order is analyzed by calculating the static structure factor in a
direction $x$ perpendicular to the dipole moments, which are aligned along
the $z$ axis. The possible existence of off-diagonal long-range order is
studied by estimating the isotropic mode $\rho_1(r)$ of the one-body
density matrix. The simultaneous existence of diagonal and off-diagonal 
long-range orders are key signatures of a supersolid state. Our work reports the
first results on $\rho_1(r)$ for the three-dimensional dipolar gas and,
with some limitations due to the finite size of the system, shows a regime
where the system can be found in a superfluid state. The range of interaction
strengths, measured in terms of $a$, where the droplets are coherent is quite
narrow, specially when compared with results derived from the extended
Gross-Pitaevskii equation. This can be partially attributed to our scaled
system, with a number of particles much smaller than the ones in typical 
experimental
realizations. However, the standard eGPE equation, incorporating the first
perturbative LHY correction, has shown relevant shortcuts when applied to
study dipolar droplets. It was shown in Ref.~\cite{fau_exp} that the critical
atom number, which is the minimum number of atoms required to form a stable
drop, is significantly underestimated within that theory. Direct calculations
using PIGS did show better agreement with the experimentally measured
critical numbers~\cite{pfau_bombin}. PIGS results point to a system that is
globally more attractive than what is predicted by the eGPE, and thus leads
to a smaller window of scattering length values where coherent drops are
observed.

In future work, we plan to simulate a similar system but in a cylindrical
geometry with periodic boundary conditions in the axial 
direction~\cite{Roccuzzo2019,Blakie2023}. In this
way, we expect to be able to evaluate the precise behavior of the one-body
density matrix at long distances, making it possible to find a reliable
estimation of the condensate fraction. On the other hand, using PIMC we are 
going to study the effect of finite temperature in the BEC to supersolid 
transition, which has shown a counterintuitive behavior where order by heating 
is observed~\cite{sohmen2021,Sanchez-Baena2023}.

\begin{acknowledgments}
We acknowledge financial support from Ministerio de Ciencia e Innovaci\'on 
MCIN/AEI/10.13039/501100011033
(Spain) under Grant No. PID2020-113565GB-C21 and
from AGAUR-Generalitat de Catalunya Grant No. 2021-SGR-01411. R.B. acknowledges 
European Union–NextGenerationEU, Spanish Ministry of Universities and Recovery, 
Transformation and   Resilience Plan, through a call from Universitat 
Polit\`ecnica de Catalunya. 
\end{acknowledgments}

\bibliography{refs} 

\end{document}